# Compositional Mapping of Martensitic and Magnetic Transitions in Ni–Fe–Ga Bulk Alloys and Microwires

L. Straka[1*], P. Veřtát[1], L. Nulandaya[1], M. Eliáš[1,3], D. Salazar[2], O. Heczko[1], R. Varga[3]

[1] FZU – Institute of Physics of the Czech Academy of Sciences, Prague, Czech Republic

[2] BCMaterials, Leioa, Spain

[3] RVactuators Ltd., Dolní Břežany, Czech Republic

* Corresponding author: ladislav.straka@fzu.cz

## Abstract

Despite extensive research on Ni–Fe–Ga magnetic shape-memory alloys, the compositional dependence of their martensitic and magnetic transition temperatures remains unclear. We analyze published bulk-alloy data alongside two complementary microwire series. While the pooled data show no unique dependence on *e/a*, elemental-concentration analysis reveals that Ni and Fe are the primary compositional controls of $M_s$ and $T_c$, respectively, within 26–28 at.% Ga: $M_s = 43.6c_{Ni} - 2079$ K and $T_c = 17.4c_{Fe} - 30$ K. At fixed Ga, this creates a trade-off: raising $M_s$ generally lowers $T_c$ and vice versa, making it difficult to keep both above room temperature. Around 25 at.% Ga, bulk data exhibit considerable scatter, partly attributable to γ-phase formation. Glass-coated microwires produced by the Taylor–Ulitovsky method, in which the γ phase has generally not been reported, extend the accessible composition range toward lower Ga contents. The $Ni_{73-x}Fe_xGa_{27}$ (x = 18–22) series and the Ga-rich $Ni_{50}Fe_{21}Ga_{29}$ and $Ni_{50}Fe_{23}Ga_{27}$ microwires broadly follow the corresponding bulk trends. Remarkably, $Ni_{50}Fe_{25}Ga_{25}$ and $Ni_{50}Fe_{27}Ga_{23}$ microwires depart sharply from these trends, combining $M_s$ at or above room temperature with $T_c > 395$ K. Their functional potential is demonstrated by approximately 10% reversible strain in $Ni_{50}Fe_{27}Ga_{23}$ microwire during thermal cycling under constant tensile stresses up to 200 MPa.

## Highlights

- Bulk Ni–Fe–Ga transition temperatures show no unique dependence on *e/a*.
- $M_s$ depends primarily on Ni and $T_c$ on Fe content, within 26–28 at.% Ga.
- Fe-rich $Ni_{50}$ microwires deviate sharply from extrapolated bulk trends.
- $Ni_{50}Fe_{27}Ga_{23}$ microwires exhibit ≈10% reversible electrothermal strain.

## 1. Introduction

Ni–Fe–Ga is an established magnetic shape memory system in which a thermoelastic martensitic transformation can be combined with useful mechanical and magnetic functionality. Conventional shape-memory and superelastic responses have been demonstrated, while magnetocrystalline anisotropy and magnetic-field-induced variant reorientation have motivated its development as a magnetic shape-memory alloy [1–5]. Particularly large magnetic-field-induced strains of approximately 8.5% were obtained by Morito and co-workers in Co-modified Ni–Fe–Ga under an assisting mechanical stress [6]. Ni–Fe–Ga has therefore long been considered as a possible counterpart to the well-known Ni–Mn–Ga magnetic shape-memory prototype, although its intrinsic magnetic-field-driven response remains more limited.

For practical use of either conventional shape-memory or magnetic shape-memory functionality, the martensitic transformation temperature and the Curie temperature are central materials-design parameters. The martensitic transformation must occur within the intended operating range, while the Curie temperature determines whether the relevant state is ferromagnetic. For thermally driven shape-memory operation, the magnetic response may additionally provide a means of sensing the structural or deformation state. For magnetic shape memory operation near room temperature, the martensitic state must be both stable and ferromagnetic within the same temperature range.

A well-established approach for modifying this balance, particularly for increasing $T_C$, is additional alloying, especially with Co. This can alter the martensitic and magnetic transition temperatures and magnetic anisotropy and has enabled substantial stress-assisted magnetic-field-induced strain [6–8]. Although quaternary optimization is important, it increases the compositional complexity. The objective of the present work is deliberately more basic: before considering further alloying, we revisit the ternary Ni–Fe–Ga system itself and ask to what extent its martensitic and magnetic transition temperatures can be rationalized directly from composition.

A substantial body of literature now reports Ni–Fe–Ga compositions together with martensitic and/or magnetic transition temperatures [1,2,9–11], providing a basis for a systematic reassessment. These data are frequently discussed in terms of the average valence-electron concentration *e/a.* However, the same *e/a* can be reached through chemically different substitution paths, which need not affect the structural and magnetic transitions in the same way. At the same time, the published values show considerable scatter. This reflects not only experimental differences but also the pronounced sensitivity of Ni–Fe–Ga to processing, atomic order, heat treatment, and phase constitution [9–12]. A further complication is the formation of the γ phase on the Ga-poor, Fe-rich side of the ternary system. In multiphase bulk alloys, γ-phase formation reduces the fraction of the transforming Heusler-derived matrix and, through elemental partitioning, can make the composition of this matrix differ from the nominal alloy composition [2,9]. Consequently, once γ-phase formation becomes significant, particularly around and below approximately 25 at.% Ga, neither nominal composition nor *e/a* calculated from it can be expected to provide a well-defined relationship with the measured transition temperatures. The large scatter in this region should therefore be regarded as a fundamental limitation of nominal-composition-based mapping, rather than solely as experimental uncertainty.

The γ phase has generally not been reported in Ni–Fe–Ga microwires prepared by the Taylor–Ulitovsky method [13–16]. Microwires may therefore provide access to low-Ga composition ranges that are difficult to realize in bulk without secondary-phase formation, thereby extending the experimentally accessible composition range for assessing composition–temperature trends.

The aim of this work is to compile published ternary Ni–Fe–Ga data relating alloy composition to the martensitic and Curie transition temperatures and to identify composition–temperature relationships useful for alloy selection and interpretation. We first examine the conventional *e/a* representation and then reorganize and analyze the data according to elemental concentrations. This analysis is complemented by Ni–Fe–Ga microwire datasets that partly overlap the bulk composition range and partly extend toward compositions that are difficult to realize in conventional bulk form. The combined analysis is used to establish the dominant compositional trends, assess their applicability to microwires, and identify where these trends remain valid or break down outside the conventional bulk composition range.

## 2. Materials and methods

The nominal compositions of the bulk alloys $Ni_{54.5}Fe_{18.5}Ga_{27}$ and $Ni_{52.8}Fe_{18.3}Ga_{26.8}Co_{2.1}$ were selected by comparison with the literature to target modulated martensite near room temperature. The Co-containing alloy was expected to combine more favorable structural and magnetic transition temperatures. The alloys were prepared by arc melting (Bühler MAM-1), followed by suction casting into rods. Temperature-dependent magnetic moment was measured at 100 Oe using a MicroSense vibrating sample magnetometer (VSM) during cooling and heating to identify the martensitic and magnetic transitions. Samples were first characterized in the as-cast state. Post-casting heat treatments were used to examine the processing sensitivity of the transformations [9,17]. Both alloys were annealed at 1473 K for 24 h and subsequently at 873 K for 90 min. The heat treatments were carried out under Ar and followed by quenching into ice water. Heat-treated samples were characterized in the same way as the as-cast samples.

Published data for ternary Ni–Fe–Ga were compiled from studies reporting alloy composition together with a martensitic transformation temperature and/or the Curie temperature [1,3,5,9–11,18–58]. In some cases, values were extracted from published plots. For consistency, $M_s$ was used as the principal descriptor of the martensitic transformation, and *e/a* was calculated as $(10c_{Ni} + 8c_{Fe} + 3c_{Ga})/100$, with concentrations in atomic percent. Structural assignments were retained as reported; the alternative 5M and 7M notations were grouped with 10M and 14M, respectively. In elemental-concentration analysis identical or near-identical compositions were grouped by rounding Ni, Fe, and Ga to the nearest 0.5 at.% and averaging the reported temperatures; each averaged composition was therefore assigned equal statistical weight. The global empirical fits were restricted to 26–28 at.% Ga. $Ga_{25}$ data were retained in the plots but excluded from the regressions because γ-phase formation and elemental partitioning make nominal composition an unreliable descriptor in this region.

The microwire data comprised experiments on the $Ni_{50}$ series, with nominal $Ni_{50}Fe_xGa_{50-x}$ compositions (x = 21, 23, 25, and 27), and published results for the complementary $Ga_{27}$ series $Ni_{73-x}Fe_xGa_{27}$ [14,59]. Glass-coated Ni–Fe–Ga microwires were produced from pre-alloyed master alloys by the Taylor–Ulitovsky method (Fig. 1); production and characterization details are given in Ref. [16]. Across the four $Ni_{50}$ compositions, the average metallic-core diameter was 11.0 µm, and the average total diameter, including the glass coating, was 32 µm [16]. The nominal core composition was taken from the X-ray-fluorescence-validated master-alloy composition. Temperature-dependent magnetic moment was measured during cooling and heating using a MicroSense VSM at 10 kOe. $M_s$ and $T_C$ were estimated from the temperature dependences of magnetic moment by the tangent-intersection method. Where the transition lay above the measurement range, only the lower bound $T_C$ > 395 K is reported.

The electrothermal response of a $Ni_{50}Fe_{27}Ga_{23}$ microwire was characterized using two in-house-built setups: a proof-of-concept demonstrator and a setup for thermal cycling under constant tensile load using electrical current. Axial strain was calculated relative to the initial gauge length as $\varepsilon = \Delta L/L_0$. The representative cycles were recorded during increasing and decreasing electrical current under nominal tensile stresses of 50, 100, and 200 MPa.

## 3. Results and discussion

### 3.1. Processing sensitivity of the bulk alloys

Figure 2 compares the temperature-dependent magnetic responses of the ternary $Ni_{54.5}Fe_{18.5}Ga_{27}$ alloy and the Co-containing $Ni_{52.8}Fe_{18.3}Ga_{26.8}Co_{2.1}$ alloy before and after heat treatment. In the as-cast state, both alloys exhibit a broad thermally hysteretic change in magnetization associated with the martensitic transformation and a sharper high-temperature decrease associated with $T_C$. The Co-containing alloy has the higher $T_C$, consistent with the known influence of Co on the magnetic transition in Ni–Fe–Ga-based alloys [8].

After heat treatment, the magnetic transition remains clearly resolved, whereas the martensitic transformation is no longer detected within the measured temperature interval in either alloy.

Magnetization alone cannot establish whether the martensitic transformation shifted outside this interval, became too weak to resolve, or was suppressed by changes in atomic order or phase constitution. The practical consequence is nevertheless the same: nominal composition does not uniquely determine the experimentally observed transformation behavior. The similar response of the ternary and Co-containing alloys also shows that this processing sensitivity is not confined to undoped Ni–Fe–Ga.

The measurements therefore illustrate an important difficulty in extracting general composition–temperature relationships from published data. Different thermal histories can produce different degrees of atomic order, matrix compositions, and phase fractions at nominally similar overall compositions [9–12,17]. Systematic compositional trends must therefore be identified against substantial processing-dependent scatter. One practical way to reduce the influence of such sample-specific variability is to average values reported for identical or near-identical compositions, as done in the present analysis.

### 3.2. Overall dataset and the limits of *e/a*

Figure 3a presents the compiled $M_s$ values in the conventional *e/a* representation and identifies the structural state reported in each source. Even within the narrow electron-count interval occupied by most ternary compositions, $M_s$ spans from well below to well above room temperature. The structural assignments are equally intermixed: 10M, 14M, mixed or complex martensites, unspecified martensite, and records without a phase-specific assignment occupy strongly overlapping regions.

The use of *e/a* in Heusler alloys is commonly rationalized by a rigid-band picture, in which alloying changes the electron count without strongly modifying the underlying density of states. Such a representation is useful for identifying broad electronic trends, and correlations with martensitic stability have been discussed for Ni–Fe–Ga [60]. In the present compilation, however, the same *e/a* can be reached by chemically different substitutions and is associated with different transition temperatures. Elemental substitution can also change bonding, magnetic exchange, atomic order, and phase constitution. We therefore retain *e/a* as a diagnostic overview but use elemental concentrations for the quantitative alloy-design analysis below.

Figure 3b focuses on 10M and 14M modulated martensites, which are of particular interest for magnetic-field-induced variant reorientation. The relatively few 10M records occur predominantly at lower Ni contents and lower $M_s$, whereas the reported 14M compositions extend toward higher Ni contents and generally higher transformation temperatures. The two regions nevertheless overlap, indicating that composition alone does not uniquely determine the modulated martensite structure.

### 3.3. Elemental composition rules in the $Ga_{26}$–$Ga_{28}$ region

#### 3.3.1. Global correlations

The pooled data become much more systematic when plotted against elemental concentrations. Figure 4a shows a strong increase of $M_s$ with Ni content, while Fig. 4b shows an increase of $T_C$ with Fe content. Symbols retain the Ga-family coding. The $Ga_{25}$ records exhibit particularly large deviations and are shown with reduced opacity. We attribute these deviations, at least partly, to γ-phase formation and associated elemental partitioning, which can cause the composition of the transforming Heusler-derived matrix to differ from the nominal alloy composition [2,9]. The $Ga_{25}$ data may therefore introduce systematic uncertainties into the composition–temperature relationships and were excluded from the regressions. Linear fits to composition-averaged $Ga_{26}$–$Ga_{28}$ data, with equal statistical weight assigned to each composition rather than to each individual literature record, yield:

$$M_s\ (\mathrm{K}) = (43.7 \pm 2.4)c_{Ni} - 2079, \quad R^2 = 0.925 \tag{1}$$

$$T_C\ (\mathrm{K}) = (17.4 \pm 1.7)c_{Fe} - 29.5, \quad R^2 = 0.835 \tag{2}$$

where $c_{Ni}$ and $c_{Fe}$ are expressed in atomic percent. The fits comprise 28 averaged compositions for $M_s$ and 24 for $T_C$, with residual standard deviations of approximately 22 K and 15 K, respectively. The $M_s$–Ni relationship is particularly strong, while the $T_C$–Fe relationship, although more scattered, remains a useful first-order predictor.

The two trends are physically reasonable. Ni enrichment at the expense of Fe has repeatedly been associated with higher martensitic transformation temperatures [1,10,60,61]. First-principles calculations identify Ni-derived electronic states and the stability difference between cubic and tetragonal structures as sensitive to composition and band filling [60,62]. By contrast, the magnetic transition is dominated by the Fe-containing exchange network. Fe carries the larger localized moment, and both direct and indirect exchange interactions, and hence the calculated $T_C$, change strongly with Fe content [60,62]. These interpretations are qualitative rather than unique microscopic proofs, but they are consistent with the distinct elemental dependencies observed in Fig. 4.

#### 3.3.2. Substitution-path dependence

Figure 5 reorganizes the composition-averaged data into two sets of substitution paths: approximately fixed-Ga families, with $M_s$ and $T_C$ plotted against Ni content (Figs. 5a,b), and approximately fixed-Ni families, with both temperatures plotted against Fe content (Figs. 5c,d). This representation allows us to examine the effects of different elemental substitutions and their relation to the global trends identified in Fig. 4.

At approximately fixed Ga, increasing Ni necessarily replaces Fe. The $M_s$ data from the $Ga_{26}$–$Ga_{28}$ families collapse onto nearly the same increasing dependence on Ni (Fig. 5a), confirming that Ni content is the dominant first-order variable, while variations in Ga content among these families have only a secondary effect. Conversely, the same substitution reduces Fe content, causing $T_C$ to decrease with increasing Ni (Fig. 5b). This reveals an important trade-off: along fixed-Ga trajectories, increasing $M_s$ through Ni enrichment simultaneously lowers $T_C$, making it difficult to achieve both transition temperatures well above room temperature.

A different behavior emerges at approximately fixed Ni, where Fe replaces Ga. The dependence of $M_s$ on Fe is relatively weak compared with its dependence on Ni and changes systematically among the fixed-Ni families (Fig. 5c): it is positive for the lower-Ni families but becomes weak or negative at higher Ni contents. In contrast, the $T_C$ data from different fixed-Ni families follow similar increasing trends with Fe content (Fig. 5d), indicating that the Fe dependence of $T_C$ is only weakly affected by Ni content within the investigated range. Thus, Fe-for-Ga substitution provides a way to increase $T_C$ without substantially reducing $M_s$, and in selected lower-Ni families, both temperatures may even increase simultaneously. This contrasts with the trade-off observed along fixed-Ga trajectories.

A two-variable regression of the $Ga_{26}$–$Ga_{28}$ data further supports the dominant roles of Ni and Fe. For $M_s$, the fitted Ni coefficient is 53.1 ± 7.0 K/at.%, whereas the Fe coefficient is 9.3 ± 6.6 K/at.% and is not statistically significant. Conversely, for $T_C$, the Fe coefficient is 19.3 ± 5.1 K/at.%, while the Ni coefficient is 2.0 ± 5.3 K/at.%. However, because Ni and Fe concentrations are strongly anticorrelated within the narrow $Ga_{26}$–$Ga_{28}$ range, these coefficients should be interpreted as local empirical trends rather than independent or universal elemental contributions.

#### 3.3.3. Practical design rules and limitations

For an initial alloy design, the one-variable relations can be inverted:

$$c_{Ni} \approx (M_s + 2079)/43.7 \qquad (3)$$

$$c_{Fe} \approx (T_C + 29.5)/17.4 \qquad (4)$$

The two target temperatures cannot be chosen independently within the calibrated composition range, because the regressions were fitted only within 26–28 at.% Ga and the three elemental concentrations must sum to 100 at.%; equivalently, $72 \leq c_{Ni} + c_{Fe} \leq 74$ at.%, with $c_{Ga} = 100 - c_{Ni} - c_{Fe}$. The two equations can therefore be used together as a simple consistency check: a selected pair of $M_s$ and $T_C$ is compatible with the fitted composition range only if the resulting Ga content remains between approximately 26 and 28 at.%. For example, 54 at.% Ni permits approximately 18–20 at.% Fe. If the calculated Ga content falls outside 26–28 at.%, the requested pair of $M_s$ and $T_C$ lies beyond the calibrated composition range and the equations should not be treated as predictions.

These equations identify promising nominal compositions, not guaranteed material states. Figure 2 demonstrates that heat treatment can shift or remove the observable transformation from the expected interval. Figure 3 shows that favorable $M_s$ and $T_C$ values do not determine the martensite type. Atomic order also changes the structural and magnetic transitions: quenching-induced reduction of $L2_1$ order can raise the martensitic transformation temperature, while increasing order tends to raise $T_C$ [17]. Candidate alloys therefore require direct verification of phase constitution, ordering state, transition temperatures, and martensite structure.

### 3.4. Microwires beyond the bulk fitting range

The empirical relations were derived in the $Ga_{26}$–$Ga_{28}$ region. Increasing Fe at fixed Ni necessarily reduces Ga and eventually enters a composition range in which bulk-alloy interpretation is complicated by γ-phase formation and elemental partitioning. Taylor–Ulitovsky microwires provide experimentally accessible trajectories through this Ga-poor region; importantly, the γ phase has generally not been reported in the microwire compositions considered here [13–16].

#### 3.4.1. Magnetic transitions in the $Ni_{50}$ series

The temperature-dependent magnetic curves of nominal $Ni_{50}Fe_xGa_{50-x}$ wires show a thermally hysteretic martensitic transformation for all four compositions (Fig. 6). For x = 21, $M_s$ is approximately 130 K and $T_C$ is about 280 K. At x = 23, $M_s$ falls to approximately 100 K while $T_C$ rises to about 390 K. A pronounced change occurs between x = 23 and x = 25: $M_s$ reaches approximately 315 K for x = 25 and 370 K for x = 27. For the latter two compositions, the Curie transition lies above the measurement limit and only $T_C > 395$ K can be stated.

The fixed-Ni series is therefore strongly non-linear. The small decrease in $M_s$ between $Fe_{21}$ and $Fe_{23}$ is followed by an increase of more than 200 K between $Fe_{23}$ and $Fe_{25}$. Thus, the Fe-rich compositions do not represent a simple continuation of the lower-Fe trend. This behavior becomes clearer below, when the wire data are compared directly with the bulk trends.

#### 3.4.2. Comparison with the bulk trends and change near $Ni_2FeGa$ stoichiometry

Figure 7 superimposes the two microwire trajectories on composition-averaged bulk reference data, with $Ga_{25}$ bulk records omitted. The $Ga_{27}$ microwire series [14,59] follows the bulk behavior closely. Increasing Ni at fixed Ga raises $M_s$, whereas increasing Fe at the expense of Ni raises $T_C$. The fixed-Ga microwires therefore reproduce the opposing trends of the structural and magnetic transitions observed in the bulk families.

The $Ni_{50}$ trajectory behaves differently. Panels 7a and 7b make this particularly evident: although the Ni content remains fixed at 50 at.%, both $M_s$ and $T_C$ vary strongly with Fe/Ga substitution. Because Ni is fixed, the bulk $M_s$ rule predicts approximately 103 K throughout the fitted range. The Ga-rich $Ni_{50}Fe_{21}Ga_{29}$ and $Ni_{50}Fe_{23}Ga_{27}$ wires, with $M_s \approx 132$ and 97 K, remain close to this expectation. The $Ni_{50}Fe_{25}Ga_{25}$ and $Ni_{50}Fe_{27}Ga_{23}$ wires depart abruptly, reaching $M_s \approx 315$ and 370 K, respectively, while retaining $T_C > 395$ K. The break therefore occurs near stoichiometric $Ni_{50}Fe_{25}Ga_{25}$ rather than as a gradual continuation of the $Ga_{26}$–$Ga_{28}$ bulk trend. $T_C$ retains the more systematic Fe dependence: it rises from about 280 K at $Fe_{21}$ to 390 K at $Fe_{23}$ and beyond the experimental limit at $Fe_{25}$ and $Fe_{27}$.

A plausible explanation is a change in Fe/Ga defect chemistry and chemical order near the $Ni_2FeGa$ composition. On the Ga-rich, Fe-deficient side, excess Ga may occupy nominal Fe-sublattice sites; on the Fe-rich, Ga-deficient side, excess Fe may occupy nominal Ga sites and produce related Fe-rich local environments. At the nominally stoichiometric $Ni_{50}Fe_{25}Ga_{25}$ composition, the average composition does not require antisite excess of either Fe or Ga, although antisite disorder, local chemical modulation, and incomplete order may still occur.

Transmission-electron microscopy of rapidly solidified $Ni_2FeGa$ has revealed interwoven Fe-rich and Ga-rich micromodulated regions formed, respectively, by Fe partially replacing Ga and Ga partially replacing Fe [63]. Mössbauer spectroscopy of a $Ni_{50}Fe_{27}Ga_{23}$ microwire found a majority magnetically ordered $L2_1$-like component together with a minority nonmagnetic Fe contribution attributed to misplaced Fe, surface environments, or antiphase boundaries [64]. Together with the strong order sensitivity reported for bulk Ni–Fe–Ga [17], these observations make a change in ordering and site occupancy across the $Ni_{50}Fe_{25}Ga_{25}$ composition a plausible explanation for the sharp rise in $M$s.

The high-$M_s$ $Fe_{25}$ and $Fe_{27}$ wires are nevertheless important independently of the microscopic explanation. They demonstrate compositions outside the bulk fitting domain in which the martensitic transformation occurs at or above room temperature while ferromagnetism persists to at least 395 K. The microwire results therefore extend the accessible composition range but cannot be regarded as a simple extrapolation of the bulk design rules.

### 3.5. Thermomechanical functionality

Figure 8 demonstrates electrothermal actuation of the $Ni_{50}Fe_{27}Ga_{23}$ microwire. Electrical current heats the wire by Joule heating, producing a visible displacement under a suspended load (Figs. 8a–c). During current cycling under constant tensile stresses of 50, 100, and 200 MPa, the wire exhibits hysteretic strain loops with a reversible strain excursion of approximately 10% (Fig. 8d). Upon reducing the current, the strain returns close to its initial value, demonstrating a substantial and reversible transformation-related response.

These measurements represent thermally driven shape-memory actuation rather than magnetic-field-induced strain, which was not investigated. They confirm that the Fe-rich $Ni_{50}Fe_{27}Ga_{23}$ microwire, whose composition lies outside the bulk fitting range, not only undergoes a martensitic transformation above room temperature but also exhibits a large recoverable mechanical strain. Its high Curie temperature ($T_c$ > 395 K) additionally provides the basis for combining such electrothermal actuation with magnetic readout or sensing.

## 4. Conclusions

The literature compilation and experimental results lead to four principal conclusions:

1. The pooled bulk Ni–Fe–Ga data show no unique relationship between valence-electron concentration *e/a* and $M_s$, $T_c$, or martensite type.

2. In contrast, plots against elemental concentrations reveal clear trends: within 26–28 at.% Ga, $M_s$ depends predominantly on Ni content and $T_c$ on Fe content, with empirical slopes of approximately 43.7 K per at.% Ni and 17.4 K per at.% Fe, respectively.
3. The $Ga_{27}$ microwires broadly follow the bulk composition trends, whereas the $Ni_{50}$ series departs sharply from their extrapolation near $Ni_{50}Fe_{25}Ga_{25}$ stoichiometry. The $Fe_{25}$ and $Fe_{27}$ microwires achieve $M_s$ at or above room temperature together with $T_c > 395$ K.
4. The $Ni_{50}Fe_{27}Ga_{23}$ microwire exhibits approximately 10% reversible strain during electrothermal cycling under tensile load, demonstrating substantial shape-memory functionality in the Ga-poor composition range.

Overall, we identify the composition range in which simple elemental trends describe the martensitic and magnetic transitions in Ni–Fe–Ga, and demonstrate that these trends cannot be simply extrapolated to Ga-poor microwire compositions near and beyond $Ni_2FeGa$ stoichiometry. This departure highlights the potential of microwire processing to access Ga-poor ternary compositions combining martensitic transformation at or above room temperature, a high Curie temperature, and large reversible thermal strain.

## Acknowledgements

The authors acknowledge funding support from the Czech Science Foundation [grant No. 24-10334S] and from the Ferroic Multifunctionalities project, supported by the Ministry of Education, Youth and Sports of the Czech Republic, Project No. CZ.02.01.01/00/22_008/0004591, co-funded by the European Union.

## Data availability

All data supporting the conclusions of this study are available from Zenodo (https://doi.org/10.5281/zenodo.23843778).

## Declaration of competing interest

The authors declare no competing interests.

## Declaration of generative AI and AI-assisted technologies in the manuscript preparation process

During the preparation of this work, the authors used OpenAI ChatGPT for language editing and to assist with the preparation and checking of plotting scripts and graphical formatting. The authors independently performed and verified the data analysis and scientific interpretation. All resulting text and figures were reviewed and edited by the authors, who take full responsibility for the content of the publication.

## FIGURES

Fig. 1. Glass-coated Ni–Fe–Ga microwires produced by the Taylor–Ulitovsky method. (a) Schematic of drawing and quenching from the molten master alloy. (b) Spool of as-produced microwire (diameter 4 cm). (c) Wire end showing the metallic core protruding slightly beyond the glass coating (SEM observation). (d) Representative micrograph of the metallic-core surface (SEM observation).

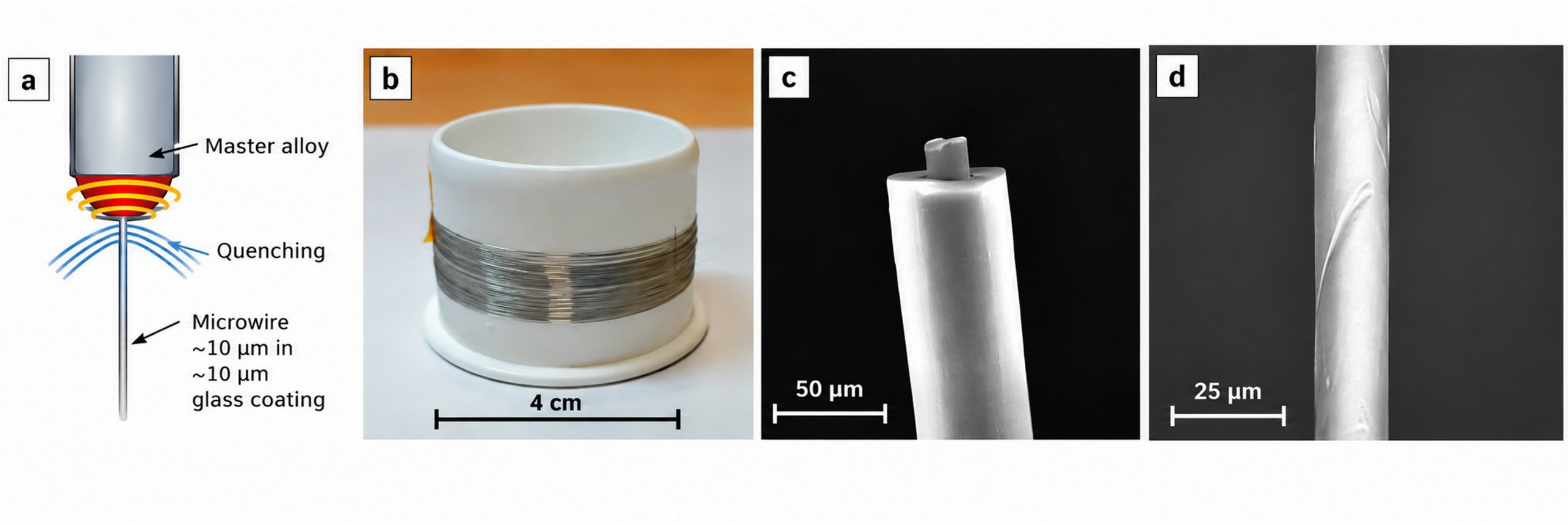

Fig. 2. Temperature-dependent magnetic moment of bulk $Ni_{54.5}Fe_{18.5}Ga_{27}$ (upper traces) and $Ni_{52.8}Fe_{18.3}Ga_{26.8}Co_{2.1}$ (lower traces) alloys in the (a) as-cast and (b) heat-treated states.

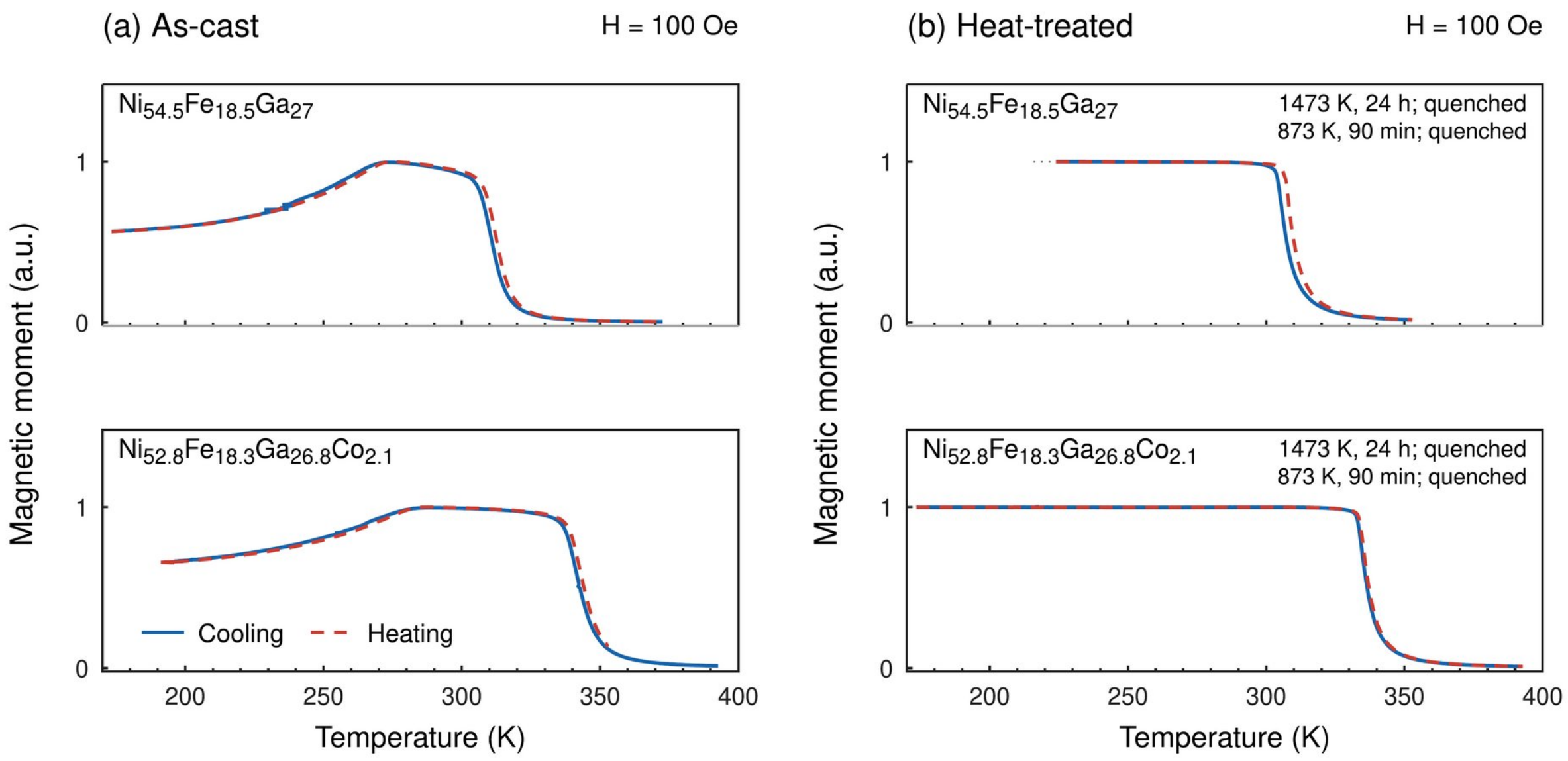

Fig. 3. Reported martensitic structures in ternary Ni–Fe–Ga alloys. (a) Martensite-start temperature $M_s$ versus valence-electron concentration *e/a*, classified by reported structural state or martensite type. (b) $M_s$ versus Ni content for records assigned to 10M or 14M martensite.

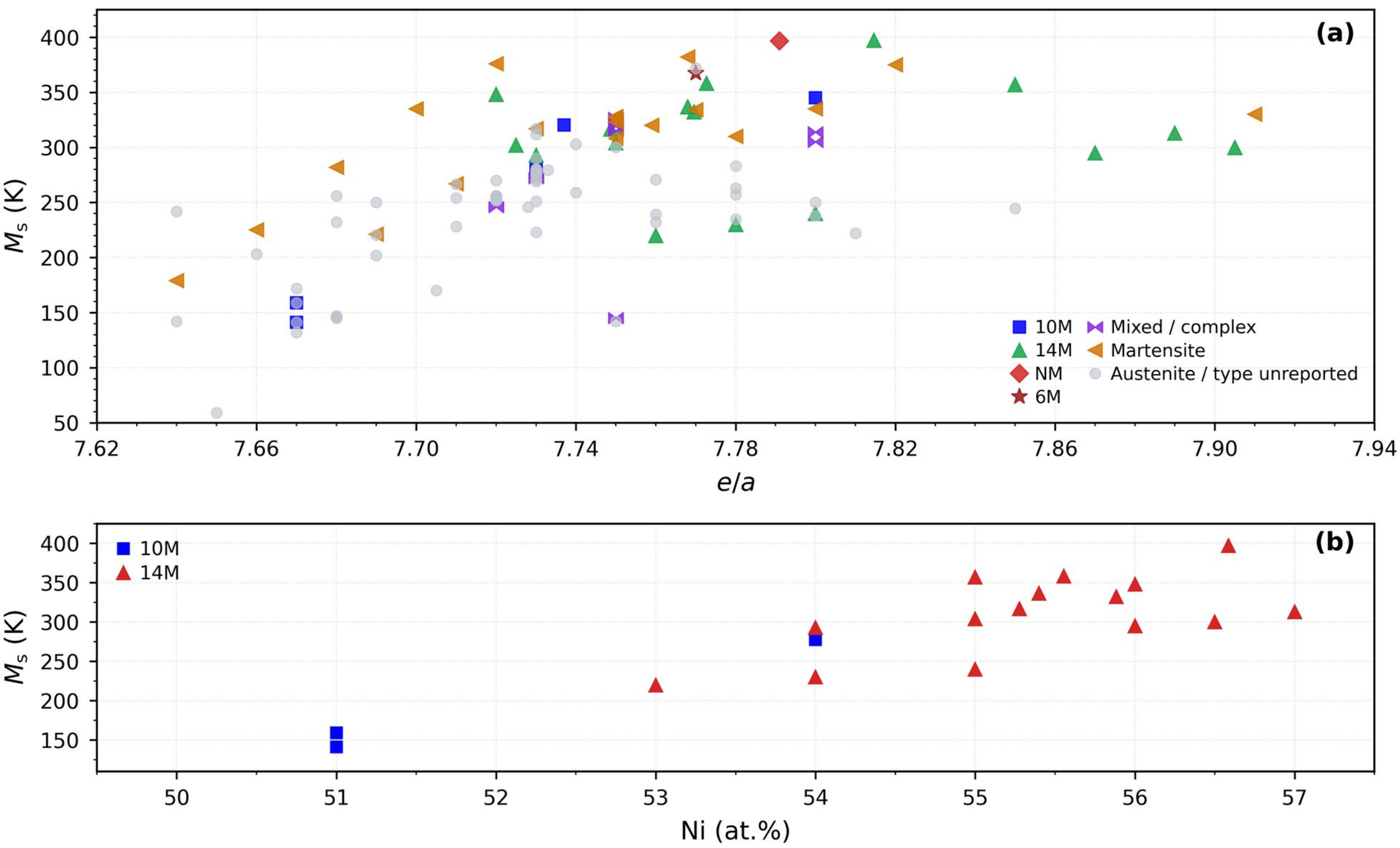

Fig. 4. Global composition trends in ternary Ni–Fe–Ga: (a) $M_s$ versus Ni content and (b) $T_C$ versus Fe content. Symbols denote Ga-content families. Lines are linear fits to composition-averaged $Ga_{26}$–$Ga_{28}$ data; $Ga_{25}$ points are shown with reduced opacity and are excluded from the fits.

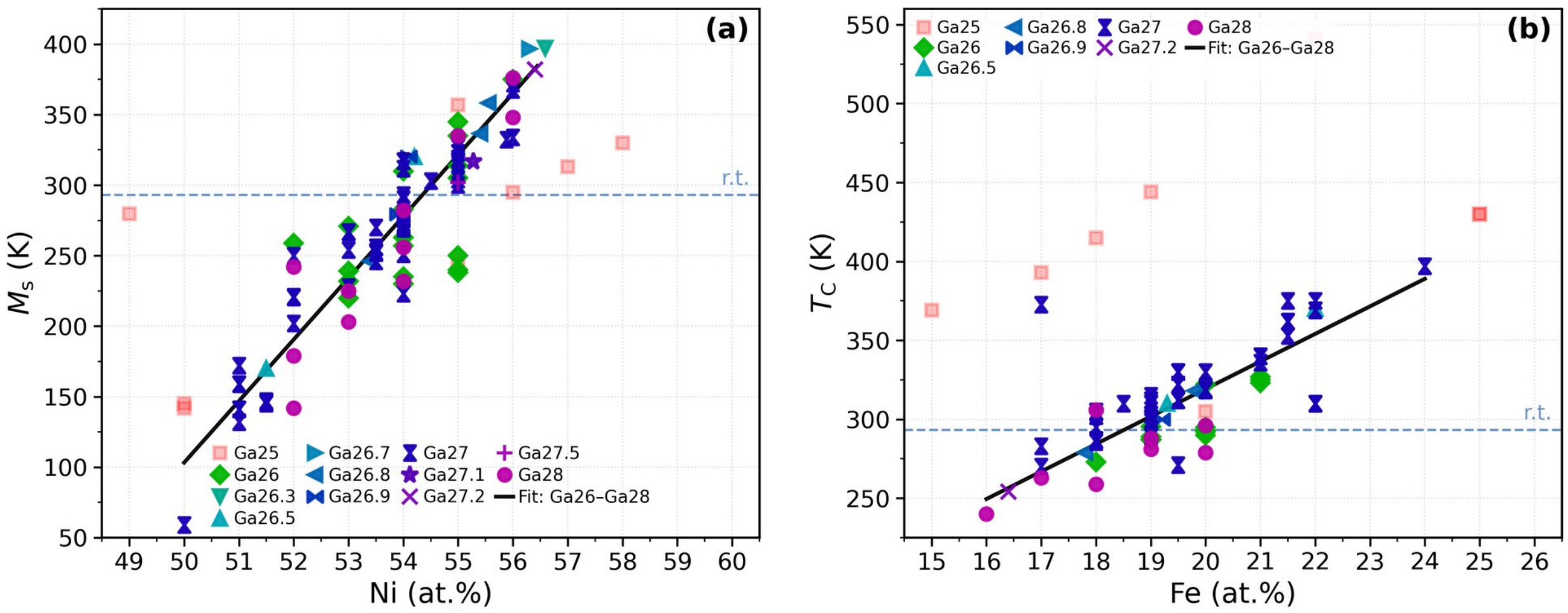

Fig. 5. Composition–temperature trends along the principal substitution paths. Lines are linear fits within the individual composition families. (a,b) $M_s$ and $T_C$ versus Ni content within approximately fixed-Ga families; (c,d) $M_s$ and $T_C$ versus Fe content within approximately fixed-Ni families. Temperatures were averaged for identical or near-identical compositions before fitting. $Ga_{25}$ points are shown with reduced opacity and are excluded from the fits.

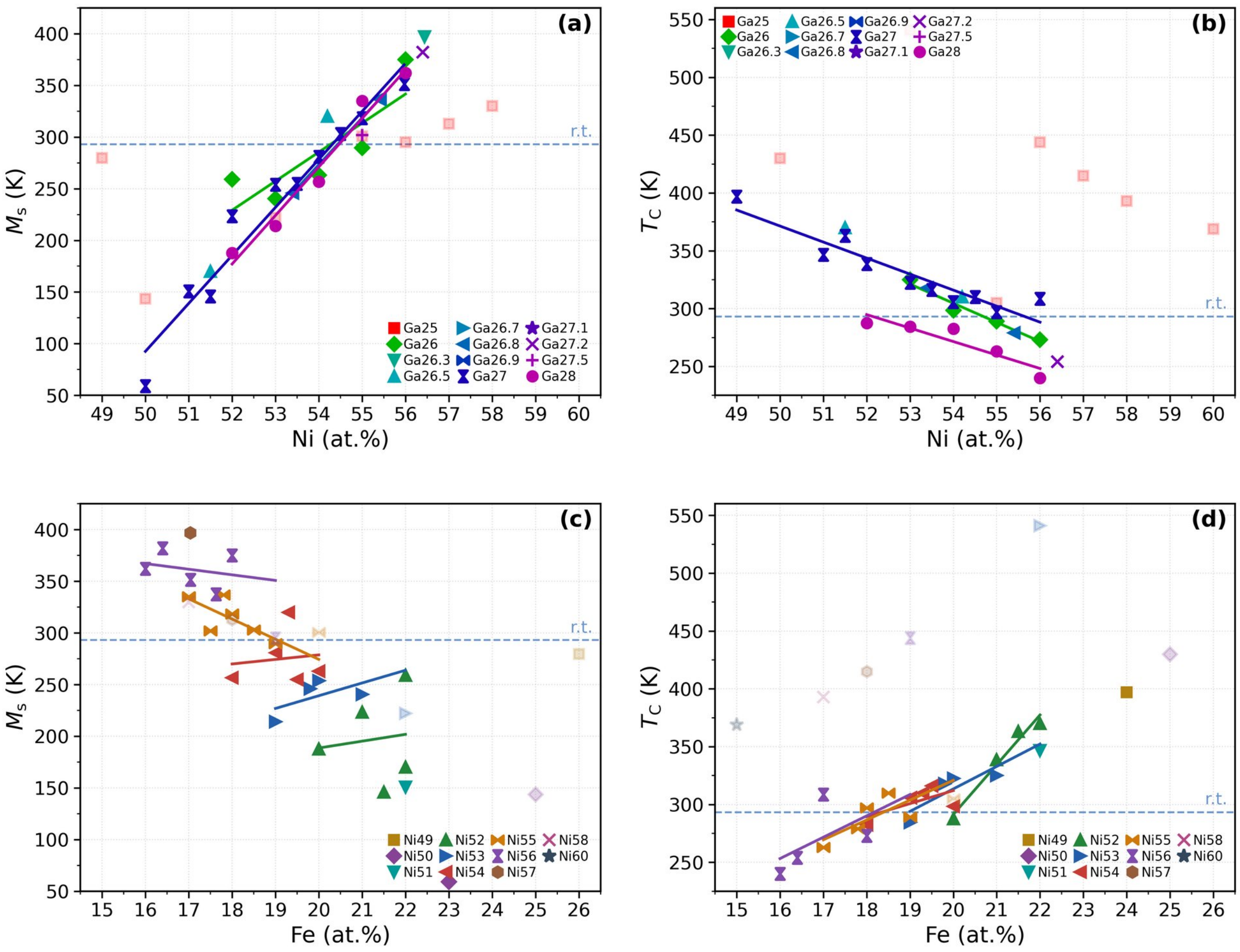

Fig. 6. Temperature-dependent magnetic moment of nominal $Ni_{50}Fe_xGa_{50-x}$ microwires (x = 21, 23, 25, and 27), measured on cooling and heating at 10 kOe. The curves are vertically offset for clarity; insets illustrate the determination of $M_s$ by the tangent-intersection method. For the $Fe_{25}$ and $Fe_{27}$ compositions, $T_C$ exceeds the measurement limit (395 K). The negative $Fe_{21}$ signal is attributed to a diamagnetic background contribution from the sample holder.

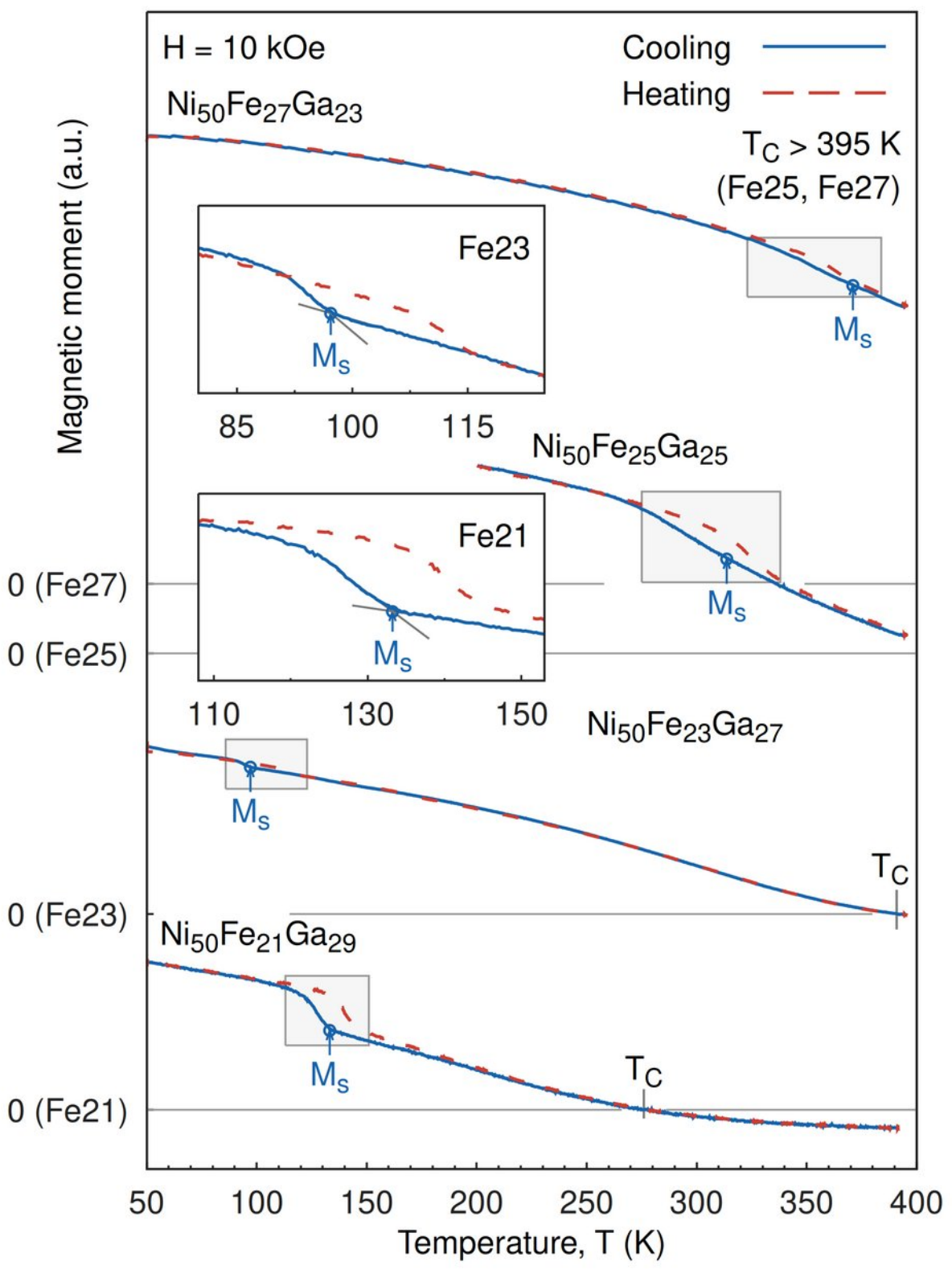

Fig. 7. $Ni_{50}$ and $Ga_{27}$ microwire series compared with composition-averaged bulk reference data. (a,c) $M_s$ and (b,d) $T_C$ plotted against Ni (a,b) or Fe (c,d) content. Bulk data are shown in grey for reference; the $Ni_{50}$ microwires [16] and $Ga_{27}$ microwires [14,59] are highlighted. Arrows indicate the lower bounds $T_C >$ 395 K.

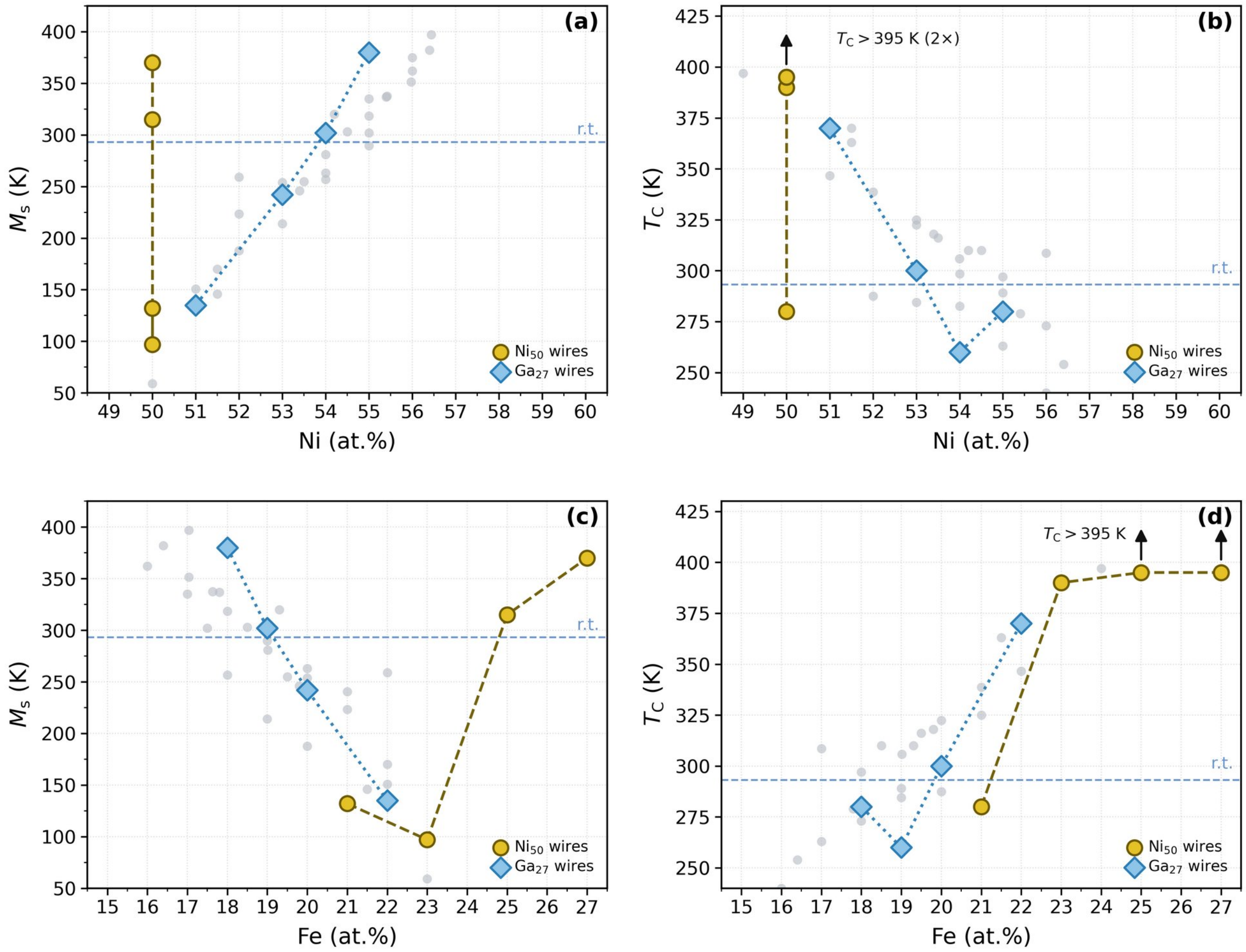

Fig. 8. Electrothermal actuation of the $Ni_{50}Fe_{27}Ga_{23}$ microwire under tensile load. (a) Current-heating configuration; (b,c) photographs of two actuation states; and (d) reversible axial strain versus electrical current during heating and cooling under nominal tensile stresses of 50, 100, and 200 MPa.

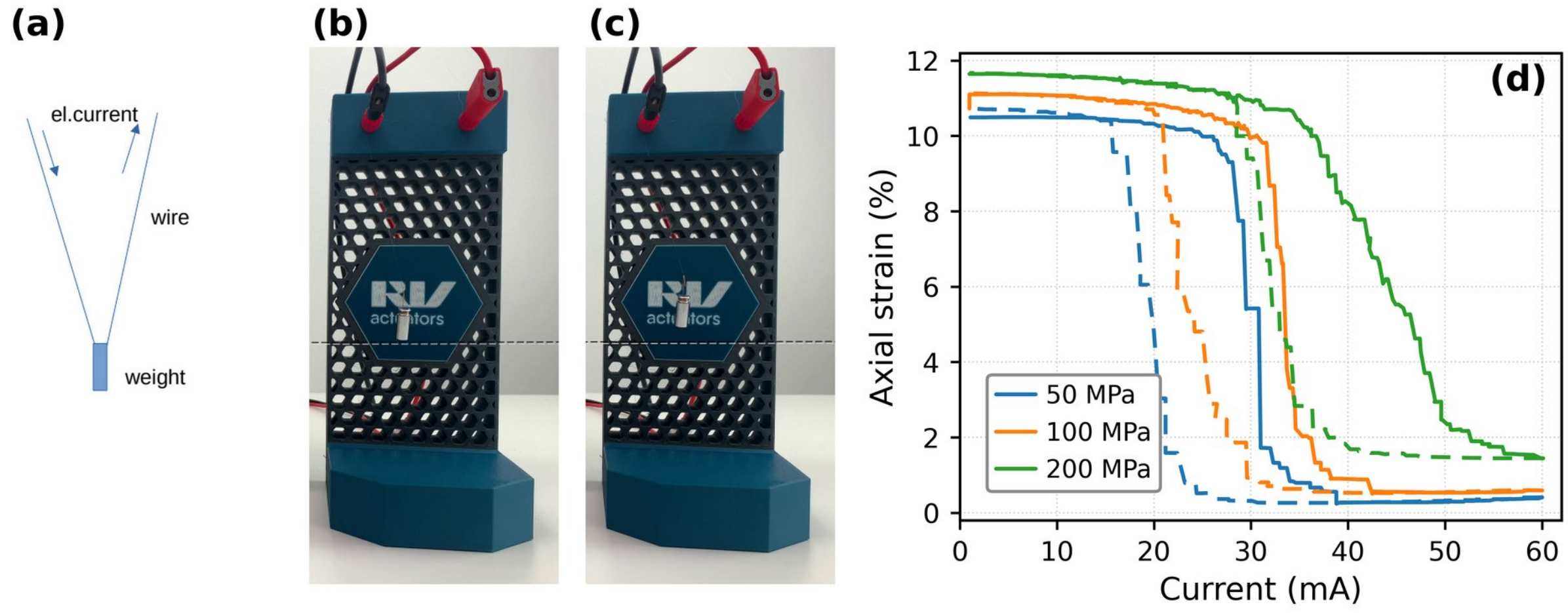